\documentclass[aps,prl,showpacs,twocolumn,groupedaddress]{revtex4-1}
\usepackage{epsfig}
\usepackage{dcolumn}
\usepackage{epstopdf}
\usepackage{graphicx}
\usepackage{ulem}
\usepackage{color}
\begin{document}
\title{Ab initio thermal conductivity of thermoelectric
 Mg$_3$Sb$_2$: evidence for dominant extrinsic effects}
\author{Maria Barbara Maccioni}
\author{Roberta Farris}
\author{Vincenzo Fiorentini}
\affiliation{Department of  Physics at University of Cagliari, and CNR-IOM, UOS Cagliari, Cittadella Universitaria, I-09042 Monserrato (CA), Italy}
\date{\today}

\begin{abstract}
The lattice thermal conductivity of the candidate thermoelectric material Mg$_3$Sb$_2$ is studied from first principles, with the inclusion of anharmonic, isotope, and boundary scattering processes, and via an accurate solution of the Boltzmann equation. We find that the anomalously low observed conductivity is due to grain-boundary scattering of phonons, whereas the purely anharmonic conductivity is an order of magnitude larger. Mass disorder due to alloying and off-stoichiometry is also found to contribute significantly to its decrease. Combining ab initio values vs sample size with measured grain-size distributions, we obtain an estimate of $\kappa$ vs T in nano-polycrystalline material  in good agreement with typical experiments, and compute the ZT figure of merit in the various cases. 
\end{abstract}
\maketitle

Thermoelectricity is emerging as a viable energy source  for a number of applications that recycle thermal waste, and materials with a high thermoelectric figure of merit 
$$ {\rm ZT}=\frac{\ \sigma S^2}{\kappa}{\rm T},$$
where $\sigma$ and $\kappa$ are the electrical and thermal conductivities, $S$ the Seebeck coefficient, and T the temperature, are currently  in growing demand. Recently, Mg$_3$Sb$_2$ has been studied fairly extensively \cite{mgsb1,mgsb2,mgsb3,mgsb4,mgsb5,sig,gb} as a prototype of 
  a family of so-called Zintl phases that have emerged as interesting candidates. The main focus has been at first on electronic properties, as the Seebeck coefficient in this family is somewhat larger than usual due to its multi-valley conduction band manifold. Equally significant for  ZT, however, is the thermal conductivity of Mg$_3$Sb$_2$ and its doped relatives, which  is unusually low for a crystalline material, namely $\kappa$$\simeq$1.5 W K$^{-1}$m$^{-1}$ around room T (with considerable experimental scatter). The thermal conductivity $\kappa$=$\kappa_{\ell}$+$\kappa_e$ is the sum of lattice and electronic  contributions; $\kappa_e$ is often modest (of order 1 W K$^{-1}$m$^{-1}$ \cite{SA}) at the typical doping used in this material and generally in thermoelectric applications, and it can be phased out by reducing the doping density. Thus,  the interesting anomaly must reside in the lattice contribution $\kappa_{\ell}$  being unusually small. 

At present, there is no theoretical estimate of $\kappa$ based  on  direct state-of-the-art calculations. In particular, it is not obvious that intrinsic vibrational properties be  responsible for the low $\kappa$. 
In this  paper we provide an ab initio assessment of lattice thermal conductivity in Mg$_3$Sb$_2$, including third-order anharmonic scattering processes, isotopic scattering, and Casimir finite-size boundary scattering. We find that the low thermal conductivity is due to microstructure size effects, i.e. to grain boundary scattering of phonons due to polycrystallinity. Boundary scattering reduces the thermal conductivity due to anharmonic scattering by as much as an order of magnitude for relevant crystallite sizes; significant isotopic scattering (peculiar to the isotopic composition of Mg and Sb) and structural anisotropy (due to the hexagonal structure) further contribute to reducing $\kappa_{\ell}$. After exploring the effects of different ingredients,  we  estimate an average $\kappa_{\ell}$ vs T in polycrystalline Mg$_3$Sb$_2$ with nanosized grains by combining ab initio values vs sample size with measured grain-size distributions. We finally use the calculated  $\kappa_{\ell}$ in a calculation of the figure of merit ZT. In all cases,  results seem  in fair to satisfactory agreement with typical experiments.

We use the Quantum Espresso suite \cite{QE} to optimize the structure of Mg$_3$Sb$_2$ and obtain the phonon spectrum, and the D3Q-Thermal2 codes \cite{D3Q,tk} for the anharmonic force constants, thermal conductivity, and $q$-dependent linewidths, including Casimir and isotopic-disorder scattering. We use 
  generalized-gradient (GGA) density-functional theory (DFT)  \cite{pbe} for electron-electron interaction, and Hartwigsen-Goedeker-Hutter \cite{hgh} 
  norm-conserving pseudopotentials for electron-ion interaction. The plane-wave cut-off is set at 50 Ry, and the $k$-points grids are 8$\times$8$\times$8 for both structure optimization and phonon dynamical-matrix calculations, 4$\times$4$\times$4 for the third-order force constants, and 10$\times$10$\times$10 for the thermal conductivity. For the thermal conductivity calculation, the "exact" iterative conjugate-gradient solution method of Ref.\cite{tk} (Sect.III) is used, with $\delta$ functions  mimicked by Gaussians with a  width of 5 cm$^{-1}$. Tests on grids, cutoff, and widths, suggest that the lattice thermal conductivity value is stable to within about 5\%.

\begin{figure}[ht]
\includegraphics[width=0.9\linewidth]{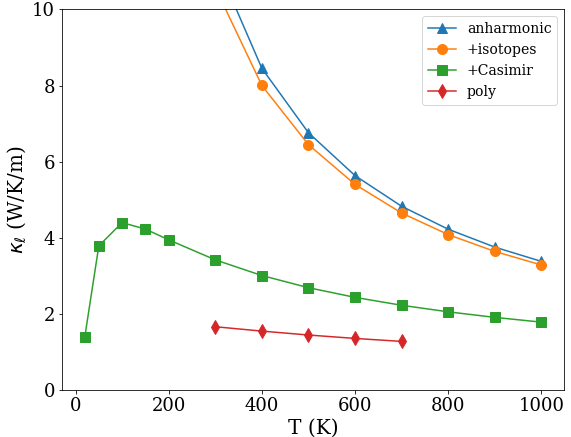}
\caption{\label{fig1} Temperature dependence of $\kappa_{\ell}$ in Mg$_3$Sb$_2$. From top: anharmonic; anharmonic and isotopes; anharmonic, isotopes and Casimir ($L$=50 nm).  The lowest curve (diamonds) is the average conductivity over an experimental  grain-size  distribution in Mg$_3$Sb$_2$ polycrystals (see text and Figure \protect\ref{fig2}).}
\end{figure}

The calculated lattice thermal conductivity $\kappa_{\ell}$ in a vibrationally-harmonic crystal is infinite. Phonon-phonon interactions due to anharmonicity cause it to become finite, with a roughly $\sim$1/T behavior above the Debye temperature, and still diverging as temperature goes to zero. If the finite size of the crystal is accounted  for \cite{redph,ziman,casimir}, $\kappa_{\ell}$ becomes finite  at zero T  and generally decreases at all T's. In addition, if different isotopes of the constituents exist, ionic mass disorder   further reduces $\kappa_{\ell}$.

Figure \ref{fig1} presents the computed lattice thermal conductivity $\kappa_{\ell}$ in several different variants. In all cases (see below for the lowest curve), we plot the inverse average of the tensor components (which are $\kappa_{\ell}^{xx}$=$\kappa_{\ell}^{yy}$$\simeq$ 1.1 $\kappa_{\ell}^{zz}$; there is no off-diagonal component in zero magnetic field). The anharmonic scattering processes result in a $\kappa_{\ell}$ (upper curve in the Figure) of over 10 at room T, which is typical of crystals, but nearly a factor 10 larger than most experimental reports. Isotopic 
disorder scattering reduces $\kappa_{\ell}$ (second curve from top), as expected  from the    significant naturally-occurring isotopic diversity of Mg and Sb \cite{nist}, but the effect is not nearly enough to cure  the discrepancy.
We then include Casimir  boundary scattering. We choose isotropic shape and, for demonstration purposes, a size of $L$=50 nm. The result is the third curve from top in Figure  \ref{fig1}, which shows that finite-size scattering at this length decreases $\kappa_{\ell}$  by a factor of about 3 at room temperature, to somewhere around a factor 2-3 the experimental value. In all calculations including Casimir scattering we have assumed the  correction factor \cite{tk} for shape and roughness \cite{redph,ziman,casimir} to be $F$=1, which we deem appropriate to isotropic grains with rough surfaces and separated by sizable disordered regions, such as those found in this material \cite{gb}. We checked, at room temperature, that the often-used \cite{tk,sp} value $F$=0.5  would in fact reinforce our conclusions, reducing $\overline{\kappa}_{\ell}$  by a further 15-20\%.

\begin{figure}[ht]
\includegraphics[width=1\linewidth]{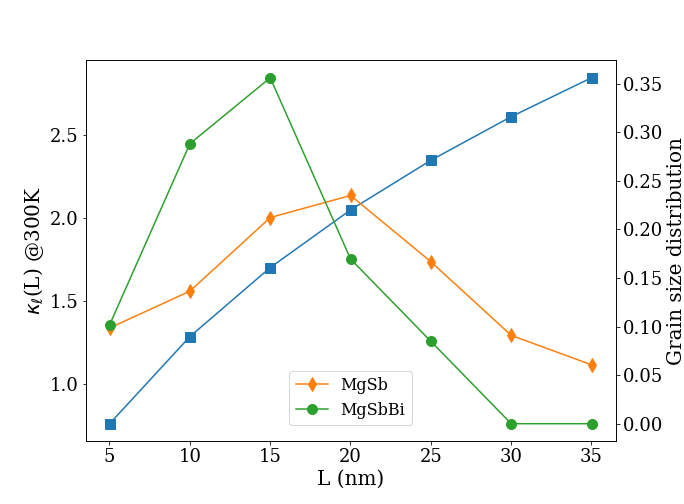}
\caption{\label{fig2}${\kappa}_{\ell}$(L)  of Mg$_3$Sb$_2$ at 300 K (squares, left vertical axis), and the distributions (from Ref.\cite{sig}, normalized; right vertical axis) of grain sizes vs $L$ in Mg$_3$Sb$_2$ and Mg$_3$Sb$_{1.8}$Bi$_{0.2}$.}
\end{figure}

Now we consider that the average value of $\kappa_{\ell}$ in  polycrystalline samples will
be determined by the grain size distribution and shape. We thus investigate the size 
dependence of $\kappa_{\ell}$, and set up a simple estimate based on actual grain 
distributions \cite{sig}. Figure \ref{fig2} reports the  size-dependent $\kappa_{\ell}$(L) 
(the usual inverse average of the components) at 300 K, along with the normalized   
grain-size distribution $n_{\rm L}$, imported from Figure 3$a$ of Ref.\onlinecite{sig} 
(the  same distribution for  a Mg$_3$Sb$_{1.8}$Bi$_{0.2}$ alloy from Figure 3$d$ of 
Ref.\onlinecite{sig}  is also reported: see below for discussion). Evidently, the grain 
distribution is quite localized over  small values (about 20 nm on average) and becomes 
negligible at larger values. To  average over the grains, we assume that thermal transport 
will occur ``in series''  across randomly-oriented  grains of size L with abundance given 
by the normalized distribution $n_{\rm L}$, so that $$\overline{\kappa}_{\ell}=1/\left[\sum_{\rm L} \frac{n_{\rm L}}{\kappa_{\ell}({\rm L})}\right].$$  
 The result at 300 K is $\overline{\kappa}_{\ell}$$\simeq$1.65, which is in line with 
 typical  experiments. (We note that the specifics of averaging are not crucial; one could 
 argue instead that $\overline{\kappa}_{\ell}$=$\kappa_{\ell}$(L$_{\rm ave}$) for an 
 average grain size, say at the peak  of the distribution, and still get essentially the same 
 value.)

\begin{figure}[ht]
\includegraphics[width=0.9\linewidth]{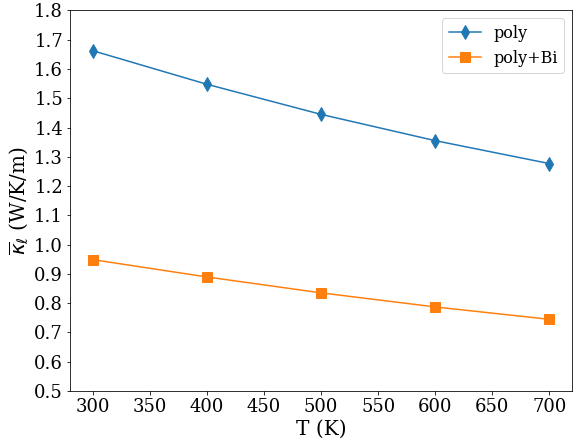}
\caption{\label{7-bi}$\overline{\kappa}_{\ell}$ in poly-Mg$_3$Sb$_2$ (diamonds, same as Figure \ref{fig1}) and Mg$_3$Sb$_{1.8}$Bi$_{0.2}$ (see text).}
\end{figure}

Repeating the L-dependent calculations at other temperatures, we finally obtain the lowest curve (diamonds) in Figure 1, which is indeed as close to the experimental data (e.g. Figure 5$c$ of Ref.\cite{sig}) as the intrinsic variability of grain sizes and shapes will reasonably allow. 
This indicates that  polycrystallinity is the likely cause of the low thermal conductivity, and is therefore an essential ingredient of thermoelectric efficiency in Mg$_3$Sb$_2$. 

In Figure \ref{7-bi} we assess qualitatively the behavior of $\overline{\kappa}_{\ell}$  in  a low-concentration Mg$_3$Sb$_{1.8}$Bi$_{0.2}$ alloy. We calculate $\kappa_{\ell}$(L)  with Bi acting as a fictitious Sb isotope of 10\% relative abundance in the isotope scattering term; in addition, we import the size distribution of  the   nanocrystalline alloy with the same composition from Figure 3d of Ref.\cite{gb} (also displayed in Figure \ref{fig2} above) and calculate the average. The two effects reduce  $\overline{\kappa}_{\ell}$ by roughly equal amounts;  both the temperature trend and the change in conductivity values are, within the limits of such simple model estimate, in fair agreement with experiments \cite{mgsb1}. The predicted conductivity would be probably further lowered if the actual phonon spectrum and anharmonic scattering in a Bi-containing alloy were included, due to the softer Bi-related modes. This calculation suggests that, while polycrystallinity remains the major factor,  mass disorder contributes significantly to the reduction of 
$\kappa_{\ell}$ (see also below).  

\begin{figure}[ht]
\includegraphics[width=0.9\linewidth]{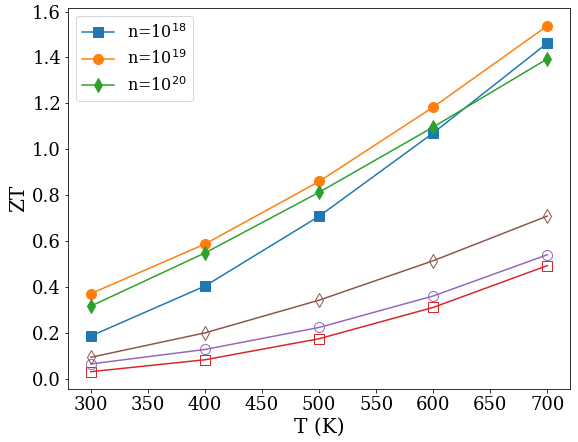}
\caption{\label{6-ZT}ZT vs T in a Mg$_3$Sb$_2$ for three $n$-type doping levels (in cm$^{-3}$)
using the perfect crystal (empty symbols) and polycrystal (filled symbols) lattice thermal conductivity.}
\end{figure}

The figure of merit ZT of Mg$_3$Sb$_2$ is reported in Figure \ref{6-ZT}, for both   the perfect crystal  and the polycrystal, for three $n$-type doping levels. Clearly, the poly and crystal situations are quite different, very possibly setting apart thermoelectrically useful vs useless material.  
ZT vs doping density has a maximum near 10$^{19}$ cm$^{-3}$ for the poly case, while it increases monotonically for the crystal case, due to the different ratio of the electrical and lattice components of $\kappa$, and the different T-behavior of the two $\kappa_{\ell}$'s (due to the very definition of ZT). Overall, ZT for our model of a $n$-doped  Mg$_3$Sb$_2$ polycrystal is essentially in the experimental ballpark (see Figure 1a, Ref.\cite{mgsb1}).

 All electronic transport coefficients  used to produce Figure \ref{6-ZT} are obtained with the Bloch-Boltzmann transport code BoltzTrap \cite{bt} including phonon and impurity scattering via an energy- and temperature-dependent relaxation time \cite{SA,libri}, which was tested in previous applications \cite{thermoCA}. The electronic structure is again obtained  within   GGA-DFT, using the projector augmented wave method as implemented in the VASP code \cite{vasp}. Full details are reported elsewhere \cite{SA}.  We use the total thermal conductivity $\kappa$=$\kappa_{\ell}$+$\kappa_e$, 
since the electronic thermal conductivity $\kappa_e$ is not generally negligible. The lattice part is either the crystal value, or the grain-average $\overline{\kappa}_{\ell}$ discussed above. For all  the electronic transport coefficients we use the crystal values, since we found that, within the model of Ref.\cite{gb}, the effects of grain boundaries on electrical quantities are marginal (of order 0.5-1\%) at our typical doping, with ZT even increasing slightly in the poly case due to a compensation of the  decrease in $\sigma$ and $\kappa_e$,  and the increase in $S$.

\begin{figure}[ht]
\includegraphics[width=0.9\linewidth]{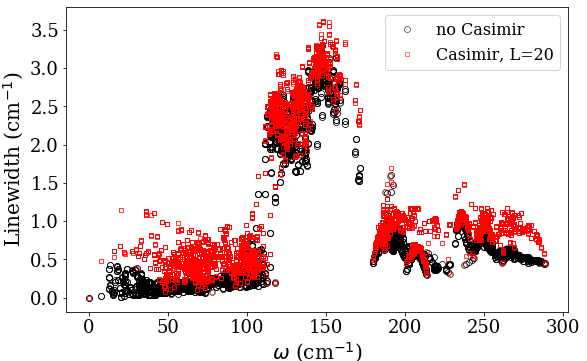}
\caption{\label{fig5}Phonon linewidths vs energy with and without Casimir scattering, $L$=20 nm, T=300 K.}
\end{figure}

We close with a brief discussion of the phonon-phonon and finite-size phonon scattering linewidths. Figure \ref{fig5}  shows linewidths vs energy for the entire spectrum. The central region between 100 and 150 cm$^{\rm -1}$ clearly dominates the linewidths; as pointed out in Ref.\onlinecite{unl}, this spectral region is the only one involving significantly the octahedrally-coordinated Mg cation, which has much longer bonds to Sb (as well as an anomalous effective dynamical Born charge  of 3.6 vs about 2 of the other two, tetrahedrally-coordinated, Mg's). 

\begin{figure}[h]
\includegraphics[width=0.9\linewidth]{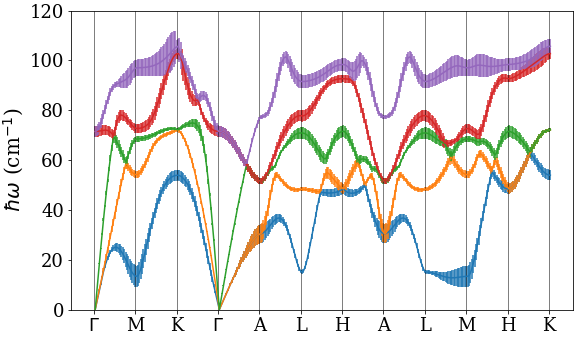}
\includegraphics[width=0.9\linewidth]{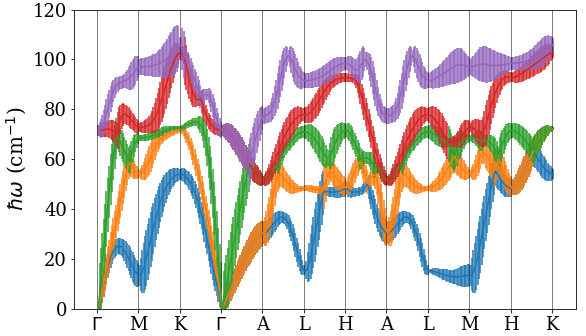}
\caption{\label{fig4} Linewidths for   low-energy phonons (including anharmonic shifts) at 300 K. Top figure: anharmonic and natural-isotope scattering, infinite size; bottom panel: same, plus Casimir scattering, $L$=20 nm.}
\end{figure}

Casimir scattering mainly affects lower-energy modes,  as borne out by  Figure \ref{fig4}, depicting the phonon dispersion  (inclusive of the quasiparticle interaction shift)  with superimposed  linewidths for the lower part of the spectrum at T=300 K. Both anharmonic and  Casimir scattering are especially significant along  the $\Gamma$-A-L-M-$\Gamma$ circuit, with the A and M points being the most significant region of scattering overall. This agrees with the identification of Ref.\cite{unl} (see in particular Figure 5 thereof) of the 	shearing transverse-acoustic mode as a significant locus of anharmonicity and scattering in this material. 

We now  discuss comparison with experiments in general and  with  other recent theoretical estimates of $\kappa_{\ell}$, referring for brevity to room T values.
Other than the present one, there appears to be no  calculation using DFPT for phonons and anharmonicity, and the full iterative Boltzmann equation solution, with inclusion of phonon depopulation and repopulation.  Two papers \cite{ref2-b,ref2-c} use  harmonic and anharmonic force constants obtained via a fit to 
frozen-phonon distortions in real-space supercells to build the relaxation time for use in 
the Boltzmann equation. Ref.\cite{ref2-b} reports $\kappa_{\ell}$$\simeq$1.5 W/(K m) computed from the Boltzmann equation in the single-mode approximation. All the force constants are obtained in supercells equivalent to a  4$\times$4$\times$4 k-grid,  very close to ours in terms of anharmonicity.  Next, Ref.\cite{ref2-c} also reports $\kappa_{\ell}$$\simeq$1.5 W/(K m) at room T, obtained again from the Boltzmann equation in an unspecified approximation; harmonic and anharmonic force constants are calculated, respectively, in  4$\times$4$\times$2 and  3$\times$3$\times$3 supercells, slightly smaller than in the other case, but probably not critical, judging from the tests provided.

 The difference with our result may stem from a combination of factors (some of which inter-related and not easily disentangled) in these technically complex calculations. First, the phonon group velocities  are calculated as finite differences on k-grids (unspecified in the papers), based on force constants obtained on relatively coarse equivalent grids (4$\times$4$\times$2 or 4$\times$4$\times$4). By contrast, our method directly evaluates group velocities from the interpolated dinamical matrix (see Ref.\cite{tk}, Sec.IV), which is obtained originally on a finer 8$\times$8$\times$6 grid.  Second, in the Boltzmann equation solution,  convergence in  k-point sampling and $\delta$-widths may be imperfect (it is unspecified in both papers)  and, third, the single-mode approximation \cite{ref2-b}  may artificially reduce $\kappa_{\ell}$ (this is a known, occasionally large, effect \cite{tk}).
   By contrast, we use the full iterative solution of Ref.\cite{tk} and carefully checked convergence in grid and widths. (From our data, we estimate a $\kappa_{\ell}$ reduction of order 30\% from single-mode solution and grid halving, but have no way of checking the effect of finite-differences differentiation; we also caution that our single-mode solution is a modified one.)
 
We also mention Ref.\cite{chino} which reports $\kappa_{\ell}$$\simeq$2.5 W/(K m), obtained via an expression involving  average Gr\"uneisen parameters and 
sound velocities.  Although a low thermal conductivity is indeed plausibly related to large Gr\"uneisen parameters (see also  Ref.\cite{unl}), this remains a simple phenomenological  model. On a different note, the estimated ZT of over 2.5  in the same paper is a significant overestimate, mainly due to of the neglect of electronic thermal conductivity, as well as to the constant-relaxation-time approximation \cite{SA}. 

As to experimental results in general, many papers report low thermal conductivities around 1 to 1.5 W/(K m) or so  in Mg$_3$Sb$_2$-based polycrystals and alloys  (including, for example, Ref.\cite{ref2-b}), and are  in general agreement with our proposed explanation.  
For single crystals, Ref.\cite{ref2-a} reports $\kappa_{\ell}$$\simeq$1.5 W/(K m) at room T in nominally pure and perfect single crystals, as well as in alloys. This low single crystal value contrasts significantly with our prediction. As to this specific work, it is unclear whether this value is representative of macroscopic single-crystals, as $\kappa$  shows a typical   finite-size downturn at low T (similar to that of Ref.\cite{mgsb3} in sintered poly material, see also below); also, it is puzzling that polycrystals and single crystals would have essentially the same $\kappa_{\ell}$.  

More generally, and aside from Ref.\cite{ref2-a} specifically, a proper comparison is predicated 
on experiments being done on disorder-free single crystals. However, single crystals are 
apparently hard to come by (and may in fact be uninteresting technologically), intentional 
alloying is ubiquitous, and general defectivity and off-stoichiometry are considerable. 
Differences in preparation (sintering, etc.),  impurity content,  texture, composition etc. cause 
significant fluctuations in the data; for example Ref. \cite{mgsb3}  gives (extrapolating in 
Fig.6c therein)  around 6   W/(K m) in $\mu$m-scale polycrystals (with L=1 $\mu$m we get about 7 W/K/m); Ref.\cite{mgsb5} reports  2.5-3 W/(K m) in hot-pressed pellets, dropping to about 1.5 W/(K m) in alloyed material. All of this suggests that Mg$_3$Sb$_2$ and related materials come in a wide spectrum of crystallite sizes and/or disorder states. 

In this context, a serious possibility that emerges is that microscopic disorder in the supposedly
single-crystal samples contributes significantly to a lowered $\kappa_{\ell}$. Indeed, thermal 
conductivity can be efficiently suppressed  by random mass disorder (an extreme example is 
$\kappa_{\ell}$=120 W/K/m in pure Si dropping by a factor 40 upon 20\% Ge random admixture 
\cite{sige});  since this scattering term is very similar to isotope disorder, we calculated $
\kappa_{\ell}$ without boundary scattering (i.e. in a periodic crystal)  with an   fictitious 
composition Mg$_{2.99}$Sb$_{1.5}$Bi$_{0.5}$ (similar to experiments in Ref.\cite{ref2-b} 
plus Mg off-stoichiometry) mimicking strong mass disorder: at 300 K we get $\kappa_{\ell}$$
\simeq$1.6 W/(K m), which is  in the experimental ballpark (this estimate of course does not 
include changes in the phonon spectrum or anharmonic couplings). So we can tentatively  
conclude that $\kappa_{\ell}$ can also be suppressed  by large mass disorder alone (as well as 
by polycrystallinity or combinations of the two). In the light of all these uncertainties, a 
definitive comparison with experiments for crystals should await  measurements in 
unambiguously large, pure, and well-ordered single crystals.
  
In summary, we proposed that the lattice thermal conductivity (and hence, by and large, the 
total conductivity in typical thermoelectric applications) of Mg$_3$Sb$_2$ is not anomalously 
small for intrinsic reasons,  but rather because of  Casimir grain-boundary scattering. While the 
bulk thermal conductivity is around 10 W/(K m) at room T, it drops to about 1.5 W/(K m)  or less for a typical  
experimental distribution of grain sizes. Strong microscopic mass disorder, 
e.g. in alloys,  also contributes significantly to the suppression of $\kappa_{\ell}$. The final 
prediction for $\kappa_{\ell}$, as well as the figure of merit ZT, in a nano-polycrystal is in good 
agreement with experiment. 

Work supported in part by UniCA, Fondazione di Sardegna, Regione Sardegna 
via  Progetto biennale 
di ateneo 2016 {\it Multiphysics approach to thermoelectricity}, and CINECA-ISCRA grants. We thank Lorenzo Paulatto for sharing his developer version of D3Q/Thermal2, and for useful discussions.

{\it Note added in proof:} 
 After acceptance, we estimated from model scattering rates \cite{SA} that in lightly doped  Mg$_3$Sb$_2$ crystals the large electron-polar phonon coupling (Fr\"olich constant$\simeq$0.6 vs. Pauling ionicity 
13\%) may  decrease the thermal conductivity by as much as 20-30\%.

\end{document}